%% file: BMCHEdges.tex
\documentclass[12pt]{iopart}
\usepackage[utf8]{inputenc}
\usepackage[english]{babel}
\usepackage{setstack}
\usepackage{amsfonts}
\usepackage{amssymb}
\usepackage{graphicx}
\usepackage{caption}
\usepackage{subfig}
\usepackage{cite}
\usepackage{color}

\begin{document}

\title[Brownian convex hulls: average number of edges]{Convex hull of $n$ planar Brownian paths:\\an exact formula for the average number of edges}
\author{Julien Randon-Furling}
\address{SAMM (EA 4543), Université Paris-1 Panthéon-Sorbonne, Centre Pierre Mend\`es-France, 90 rue de Tolbiac, 75013 Paris, France}
\ead{Julien.Randon-Furling@univ-paris1.fr}

\begin{abstract}
We establish an exact formula for the average number of edges appearing on the boundary of the global convex hull of $n$ independent Brownian paths in the plane. This requires the introduction of a counting criterion which amounts to ``cutting off'' edges that are, in a specific sense, small. The main argument consists in a mapping between planar Brownian convex hulls and configurations of constrained, independent linear Brownian motions. This new formula is confirmed by retrieving an existing exact result on the average perimeter of the boundary of Brownian convex hulls in the plane.
\end{abstract}
\submitto{JPA}

\section*{Introduction}

\begin{figure}[h]
\begin{center}
\input{Fig1.tex} 
\caption{A realization of $\partial \mathcal{C}_9$ (solid black line), the boundary of the convex hull of $n=9$ independent planar Brownian paths, all starting from the origin and having the same duration. (The dashed lines are canonical $x$ and $y$-axes.) Each planar Brownian path is defined as having for coordinates two independent linear Brownian motions.}\label{Fig1}
\end{center}
\end{figure}
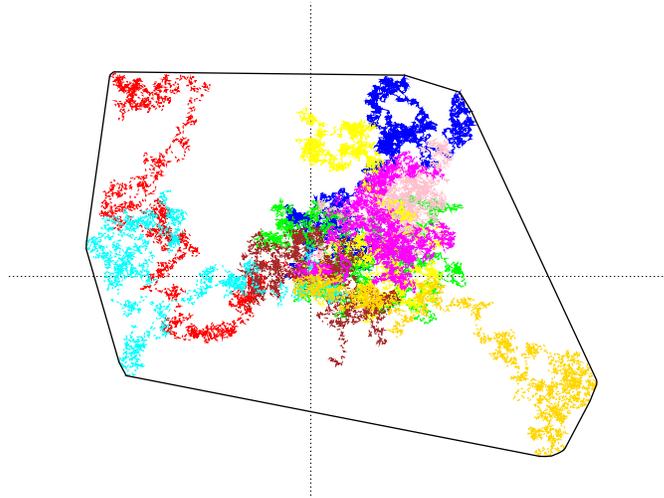

Consider $n$ independent Brownian paths in the plane, all of the same given duration and starting from the same point, and let $\mathcal{C}_n$ denote their global convex hull, that is the smallest convex set containing all $n$ paths (\fref{Fig1}). As an example of randomly-produced convex shapes based on stochastic processes, $\mathcal{C}_n$ (particularly when $n=1$) has received continuous attention over the past decades, in the fields of probability theory \cite{Lev,Kin1,Kin2,Le1,Ta,Ev,El,KBSM,CHM,Kho,Verz,BiLe} and stochastic geometry \cite{Go, Go2}, with applications ranging from image processing and pattern recognition \cite{CHIm} to statistical physics \cite{SMACRF}, ecology \cite{Worton, RFSMAC} and biophysics \cite{DNAWP}. (We focus on Brownian convex hulls, given the fundamental position that Brownian motion has been occupying in the mathematical, physical and biological sciences for more than a century \cite{AEinst,AEinst2,Kac,UhlOr,UhlOr2,Dupl,FrKr}; the literature on random polytopes is vast, and it is not our aim here to present even just an overview of it: we refer the reader to \cite{B2,WW,RSRev,Bar,BaRe} for surveys.)

Applications along with theoretical considerations lead to common, typical questions about random polytopes, pertaining for instance to the average values of some elementary geometrical quantities such as: perimeter, enclosed area or volume, number of edges or sides. In the case of the global convex hull of $n$ planar Brownian paths, exact formulae for the average perimeter and enclosed area exist in the literature: see \cite{SW,El,Ta} for $n=1$ and \cite{El, RFSMAC} for $n\geq2$. The $n\geq2$~-~case led to the design of a general method for the computation of average perimeters and areas, shedding light on the links between random convex hulls and extreme-value statistics \cite{SMACRF}. However, somewhat finer aspects concerning the shape of these hulls do not seem amenable to this method. For instance, the intuitive conjecture that, for large $n$ but a fixed, finite common duration, $\mathcal{C}_n$ tends in some sense to a circular shape has been proven recently using other techniques \cite{DaYu}. We aim to pursue here the study of the shape of  $\mathcal{C}_n$ by examining the set of line segments which compose its boundary $\partial \mathcal{C}_n$: indeed, if Brownian convex hulls do not have any isolated extreme point \cite{Ev}, they consist almost surely of countably many line segments \cite{El, Ev, CHM}. This suggests that, provided a regularization is introduced to ``cut off'' some of the ``smaller'' edges, one can compute the average number of edges on the convex hull of planar Brownian paths.

In the mathematical literature, exact formulae for average numbers of edges, sides or vertices are to be found in contexts where the random convex hull is that of a discrete set of independent random points \cite{RS,Ef,Ca,Ray2,EG,Bro,Ald,Cal}. But similar results remain rare in the case of dependent points correlated by their being the successive positions of a random walk. Our starting point is one of these results: a combinatorial lemma by G. Baxter \cite{Bax, SnS} leading to an exact formula for the average number $\langle V(N)\rangle$ of edges on the convex hull of an $N$-step random walk in the plane:
\begin{equation}
\langle V(N)\rangle = 2 \sum_{k=1}^{N}\frac{1}{k}. \label{BF}
\end{equation}

In this paper, we first show that, with a proper regularization, Baxter's formula extends to the convex hull of a single planar Brownian path precisely as one would expect by taking the continuous limit:
\begin{equation}
\langle V_{1}(T,\Delta t)\rangle = 2 \ln\left(\frac{T}{\Delta t} \right),
\end{equation}
where $T$ is the duration of the Brownian path, and $\Delta t$ is a regularization factor representing the minimum ``Brownian-path time'' along edges for them to be counted (see section~\ref{case1}).

We then derive our main result, an exact formula for the average number of edges on the boundary of the convex hull of $n$ independent Brownian paths of common duration~$T$, starting at the same point (section~\ref{casegen}):
\begin{equation}
\langle V_n(T,\Delta t)\rangle= \alpha _n(T,\Delta t) + \beta _n(T,\Delta t), \label{MF}
\end{equation}
where $\alpha_n(T,\Delta t)$ and $\beta _n(T,\Delta t)$ can be expressed as triple integrals.

Finally, as a confirmation and illustration of our main formula~\eref{MF}, we show how the underlying probability distribution allows one to retrieve the existing formula for the average perimeter of the global convex hull of $n$ independent Brownian paths (section~\ref{cper}).

\section{The $n=1$~-~case}
\label{case1}
For $T>0$ and $t\in [0,T]$, let $B(t)=\left(u(t),v(t)\right)$ denote the position of a two-dimensional Brownian motion at time $t$, that is: the coordinates $u(t)$ and $v(t)$ of $B$ in the canonical basis of $\mathbb{R}^2$ are independent, standard one-dimensional Brownian motions. They satisfy the simple Langevin equations:
\begin{equation}
\frac{du}{dt}=\eta_u (t),\qquad \frac{dv}{dt}=\eta_v (t),\nonumber
\end{equation}
where $\eta_u, \eta_v$ are independent copies of Gaussian white noise with null average and $\delta$-correlation, ie for all $t,t' \in [0,T]$:
\begin{equation}
\left\lbrace
\begin{array}{l}
\langle \eta_u(t)\rangle = \langle \eta_v(t)\rangle = 0\\
\langle \eta_u(t)\eta_u(t')\rangle =\langle \eta_v(t)\eta_v(t')\rangle=\delta (t-t')\\
\langle \eta_u(t)\eta_v(t')\rangle = 0
\end{array}
\right., \nonumber
\end{equation}
$\delta$ being Dirac's delta function, and $\langle\dots\rangle$ standing for the average over all realizations. We set $B(0)=(0,0)$ and write $\Gamma _B=B([0,T])$ for the Brownian path and $\partial \mathcal{C}_1$ for the boundary of its convex hull.
\medskip

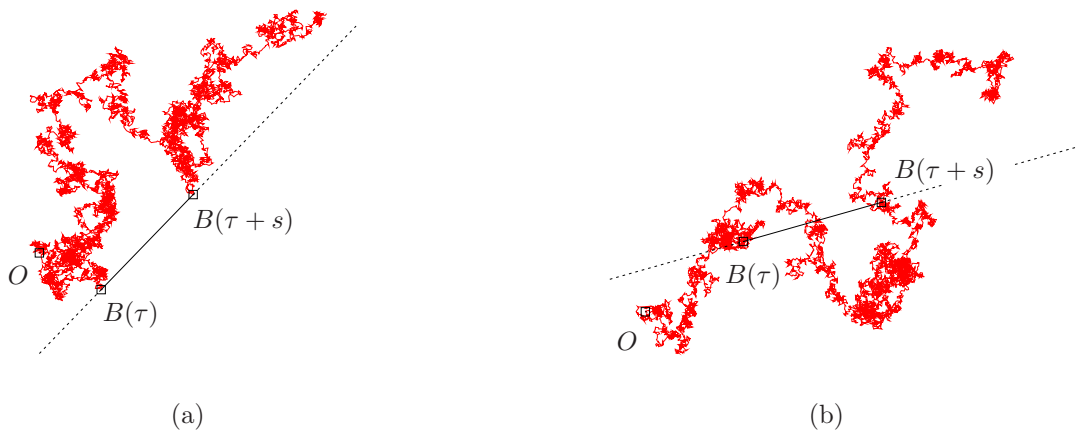
\begin{figure} 
\subfloat[]{ 
\label{fig2a} 
\begin{minipage}[b]{0.45\linewidth} 
\centering \input{Fig2a.tex}
\end{minipage}} 
\hfill 
\subfloat[]{ 
\label{fig2b}
\begin{minipage}[b]{0.45\linewidth} 
\centering \input{Fig2b.tex}
\end{minipage}} 
\caption{Given a Brownian path $\Gamma_B$ of duration $T$ starting from the origin $O$, and given times $s\in [\Delta t,T]$ and $\tau \in [0,T-s]$, $\Gamma_B$ may be found to: (a) lie on one side only of the line through $B(\tau)$ and $B(\tau+s)$; or (b) cross it. In case (a), the line segment (solid black) joining $B(\tau)$ and $B(\tau+s)$ shall belong to the boundary of the path's convex hull ; in case (b), it shall not.} 
\label{Fig2}
\end{figure}

Baxter's derivation of formula (\ref{BF}) relies on a combinatoric lemma that can be interpreted as a computation of the probability that the line segment joining two points of the walk separated by $k$ steps belongs to the boundary of the walk's convex hull. Similarly, in the case of a single Brownian path, we pick $s$ and $\tau$, such that $\tau \in [0,T]$ and $(\tau + s)\in [0,T]$, and we wish to compute the probability for the line segment joining $B(\tau)$ and $B(\tau + s)$ to belong to~$\partial\mathcal{C}_1$ (note that for $s>0$, one has almost surely $B(\tau) \neq B(\tau + s)$). Among all possible realizations of a planar Brownian path starting at the origin and having duration $T$, some will indeed be such that the line segment $[B(\tau)B(\tau+s)]$ appears on the boundary of the path's convex hull. Let us call such realizations $(s,\tau)$-paths. Using the Wiener measure for planar Brownian paths, we wish to compute the measure of the subset of $(s,\tau)$-paths, that is, the proportion of such paths among all possible realizations. Before we start this computation proper, let us observe that $s$ represents the ``Brownian-path time'' between the two endpoints of the line segment under consideration. This time interval can be used to regularize the number of edges by requiring that:
$$s\in [\Delta t,T],$$
where $\Delta t$ is some small, positive constant allowing one to actually ``cut off" in our counting those segments that join points reached within too small a time lapse by the Brownian particle.\\
Thus, once the parameter $\Delta t$ has been chosen, our question becomes: for fixed $s\in [\Delta t,T]$ and $\tau \in [0,T-s]$, what is the probability that the whole path $\Gamma_B$ lies on one side only of the line through $B(\tau)$ and $B(\tau+s)$, ie what is the probability of having \fref{fig2a} rather than \fref{fig2b}?

Let us examine $\Gamma_B$ in the coordinate system shown in \fref{fig3a}: the line through $B(\tau)$ and $B(\tau+s)$ defines the $x$-axis, and the line perpendicular to it and through $B(0)$ defines the $y$-axis. Note that if $\tau > 0$, $B(0)=y_0 \neq 0$ almost surely, so one can fix the orientation of the axes by choosing $B(0)$ to indicate the positive or negative side of the $y$-axis. By property of isotropic planar Brownian motion, the new coordinates $x(t)$ and $y(t)$ of the path remain independent one-dimensional Brownian motions. We shall call this coordinate system the $(s,\tau)$-adapted coordinate system, because of the correspondence between planar Brownian paths for which the line segment $[B(\tau)B(\tau+s)]$ belongs to the boundary of the path's convex hull, and pairs of linear, standard Brownian motions $(x,y)$, where $x$ is unconstrained except for $x(0)=0$, while $y$ satisfies the following constraints, as illustrated in~\fref{fig3b}:
\begin{enumerate}
\item[C1.] $y$ starts at $y_0>0$ (say) and hits $0$ for the first time at $t=\tau$;
\item[C2.] $y$ hits $0$ again at $t=\tau +s$ but remains strictly positive between $t=\tau$ and $t=\tau +s$;
\item[C3.] $y$ does not hit $0$ after time $t=\tau + s$, it remains strictly positive up to time $t=T$.
\end{enumerate}
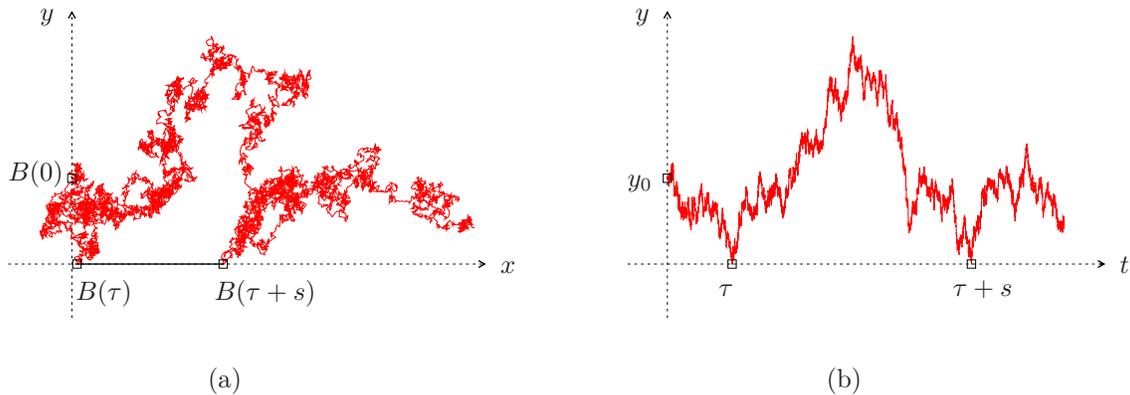
\begin{figure}
\subfloat[]{ 
\label{fig3a} 
\begin{minipage}[b]{0.45\linewidth} 
\centering \input{Fig3a.tex}
\end{minipage}} 
\hspace{0.9cm}
\subfloat[]{ 
\label{fig3b}
\begin{minipage}[b]{0.45\linewidth} 
\centering \input{Fig3b.tex}
\end{minipage}} 
\caption{The $(s,\tau)$-adapted coordinate system (a):  the line through $B(\tau)$ and $B(\tau+s)$ defines the $x$-axis, and the line perpendicular to it and through $B(0)$ defines the $y$-axis. If $[B(\tau)B(\tau+s)]$ belongs to the boundary of the path's convex hull, the $y$-coordinate will be a combination of Brownian meanders and excursions (b).} 
\label{Fig3}
\end{figure} 

Now, reasoning in the other direction, one can obtain a planar Brownian path $\Gamma _B$ for which the line segment $[B(\tau)B(\tau+s)]$ belongs to the boundary of the path's convex hull by picking two independent linear Brownian motions $x$ and $y$ behaving as stated above. Since the probability measure for planar paths is just the product measure derived from the measure for linear paths, one can compute the probability content of the subset of $(s,\tau)$-paths as the product of the probability contents of the corresponding $x$ and $y$ motions. 

The factor corresponding to the $x$-part is simple: any $x$-motion satisfying $x(0)=0$ is acceptable, so this will contribute a factor of $1$ in the joint probability measure.

As for the factor corresponding to the $y$-motion, it can be decomposed as the product of three terms thanks to the Markovian nature of Brownian motion ($s$ and $\tau$ being fixed):
\begin{enumerate}
\item the probability that the first part of the $y$-motion has a first-passage time equal to~$\tau$ (or equivalently, in reverse time, that it is a Brownian meander of duration $\tau$);
\item the probability that the second part is a Brownian excursion of duration $s$;
\item the probability that the third part is a Brownian meander of duration $T-(\tau + s)$.
\end{enumerate}
(Meander and excursion refer to the usual terminology for various types of constrained one-dimensional Brownian motion \cite{KLC1,KLC2, Finchvariants}.) \\
The first and third of these terms are readily found in the literature, since both can be viewed as the probability that a Brownian motion starting at some positive point first hits $0$ at a given time \cite{Feller,KLC2,SM,Red}. Following \cite{KLC2}, one can also write the second term from first-passage time densities. Indeed, if one introduces an intermediate time $\kappa \in [\tau, \tau+s]$, the probability that $y$ is an excursion between $\tau$ and $\tau+s$ can be split at $\kappa$:  after time~$\kappa$, $y$ should first hit $0$ at time $\tau+s$; before time $\kappa$, $y$ should have last hit $0$ at time $\tau$ (this is a last-exit time, as in \cite{DMSY,ACJDProp}. Conditioned on $y(\kappa)=r>0$, this is equivalent to picking two Brownian motions starting at $r$, requiring that they have first-passage times $\kappa-\tau$ and $\tau+s-\kappa$, and then putting them together back to back. Conveniently, the intermediate time $\kappa$ vanishes when one integrates over the intermediate position $r=y(\kappa)$.

The threefold probability for the $y$-motion should be carefully normalized by that of a standard one-dimensional Brownian motion $z$ subject not to the same constraints of course but to the same conditioning as $y$: $z$ should be ``pinned'' at the same points $z(\tau)=z(\tau +s)=0$, $z(0)$ should be positive, and, most important, $z(\kappa)$ should be positive too.
\medskip

We are now in a position to write $f(s,\tau)$, the measure of the subset of $(s,\tau)$-paths, in terms of $p(t;u,v)$ and $g(t;0,v)$, where:
\begin{equation}
p(t;u,v)=\frac{1}{\sqrt{2\pi t}}\exp\left[ -(v-u)^2/2t\right] \label{fprop}
\end{equation}
is the standard Wiener measure for a Brownian motion starting at $u$ and reaching $v$ at time $t$ (this density is seen physically as the propagator of a free Brownian particle on a line); and:
\begin{equation}
g(t;0,v)=\frac{v}{\sqrt{2\pi t^3}}\exp\left[ -v^2/2t\right] \label{mprop}
\end{equation}
is the measure of a Brownian motion starting at $0$ and reaching $v>0$ at time $t$ without re-hitting $0$ in-between (see \textit{eg} \cite{KLC2}: this is the same as the first-passage time density for a Brownian motion starting at $u>0$, and $\left[g(t ; 0,u)\, /\int_0^\infty g(t ; 0,v)\, \rmd v \right]$ is indeed the propagator for the Brownian meander).
\bigskip

Inserting the normalization detailed above and integrating over the initial, intermediate and final positions, one obtains:
\begin{eqnarray}
\fl f(s,\tau)= 2 \times \frac{\int_0^\infty\,g(\tau ; 0,y_0)\,\rmd y_0 }{\int_0^\infty p(\tau ; z_0,0)\, \rmd z_0}\ \times \ \frac{\int_0^\infty \, g(\kappa-\tau ;0,r)\,g(\tau+s-\kappa ; 0,r)\,\rmd r}{\int_0^\infty \, p(\kappa-\tau ; 0,r')\,p(\tau +s-\kappa ; r',0)\,\rmd r'}\ \times \nonumber\\
\frac{\int_0^\infty \, g(T-(\tau +s);0,y_T)\,\rmd y_T}{\int_{-\infty}^\infty \, p(T-(\tau +s); 0,z_T)\, \rmd z_T}.\label{1Dens}
\end{eqnarray}
Note that we have emphasized in the writing of the normalization the fact that it corresponds to a motion subject to the same conditioning as $y$: hence the different ranges of integration in the three parts of the normalization. Note also that we set arbitrarily $y_0 > 0$ in the constraints on the $y$-motion, but there are two sides to the line through $B(\tau)$ and $B(\tau+s)$, contributing an overall factor of $2$ when moving from the probability measure of linear realizations to that of planar realizations. 

Incidentally, the middle term in \eref{1Dens} can be interpreted as the probability that a planar Brownian path of duration $s$ lies on one side only of the line joining its endpoints: this is $2\, \rmd s\, /s$, which is to be compared to $2/k$, the probability that a discrete-time random walk formed with $k\geq 2$ steps lies on one side only of the line joining its initial and final positions \cite{Bax}.
\medskip

Now, performing the integrations in \eref{1Dens} yields:
\begin{equation}
f(s,\tau)=\frac{2}{\pi s\sqrt{\tau[(T-s)-\tau]}}.\label{f}
\end{equation}
And finally, $f(s,\tau)\rmd s\,\rmd \tau$ represents the probability for $[B(\tau)B(\tau+s)]$ to belong to $\partial C_1$. Therefore, summing over all possible line segments with~$s\geq\Delta t$:
\begin{eqnarray}
\langle V_{1}(T,\Delta t)\rangle &=& \int_{\Delta t}^T \int_0^{T-s}\,f(s,\tau)\,\rmd \tau \,\rmd s\nonumber \\
&=&2 \int_{\Delta t}^T\,\frac{\rmd s}{s}\nonumber \\
&=& 2 \ln\left(\frac{T}{\Delta t} \right),\label{Fn1}
\end{eqnarray}
as expected by taking the continuous limit of Baxter's formula (\ref{BF}).
\medskip

Before we proceed to the general case, let us remark that \eref{f} can be interpreted as the product of: (i) a term corresponding to the probability that a Brownian motion of duration $s$ is an excursion and (ii) a time-of-maximum density. This can be understood as follows: if one discards the excursion of duration $s$ between $\tau$ and $\tau+s$ and joins the two parts left, one obtains a Brownian motion of duration $T-s$ constrained to remain positive except at time $\tau$ where it hits $0$. Conditioned on its existence, the distribution of this time $\tau$ is the same as that of the time at which a Brownian motion of the same duration attains its maximum --- which is given by Lévy's famous arcsine law \cite{Lev}, $f(\tau)=1/\left[\pi\sqrt{\tau[(T-s)-\tau]}\right]$.

\section{The general case}
\label{casegen}
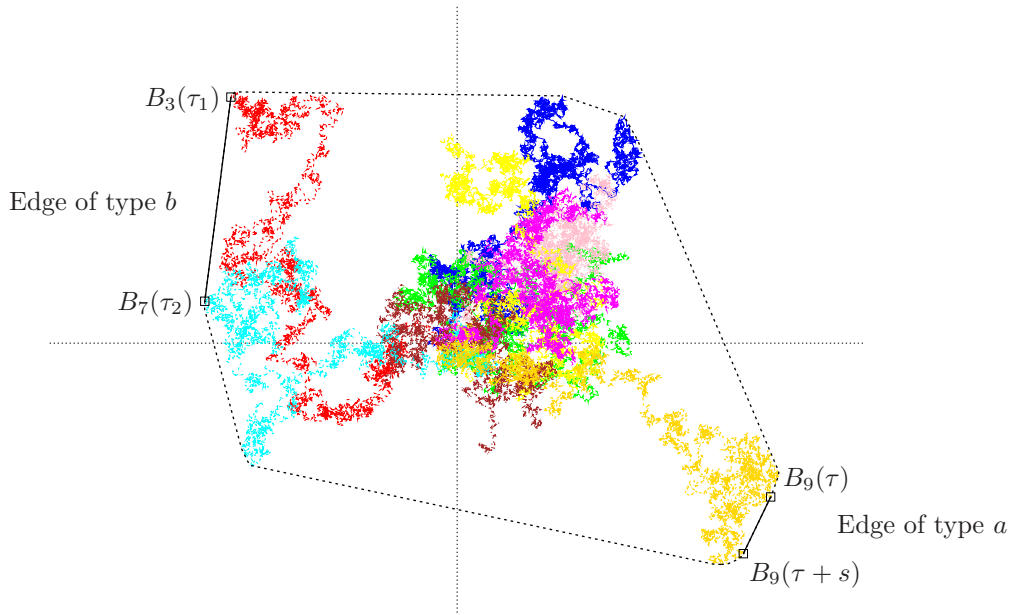
\begin{figure}[h]
\begin{center}
\input{Fig4.tex} 
\caption{A sample realization of the convex hull of $n=9$ Brownian paths in the plane, as in \fref{Fig1}. Edges on the boundary (dotted black line) of the convex hull can be of two types: ($a$) joining two points belonging to one of the paths (here for instance two points on $B_9$ (solid black line in the bottom right-hand corner)), or ($b$) joining two points belonging to two distinct paths (here on $B_3$ and $B_7$ (solid black line in the top left-hand corner)).}\label{FigEdges}
\end{center}
\end{figure}

We now consider the global convex hull of $n\geq2$~independent planar Brownian paths $B_1, B_2,\dots,B_n$, all starting at the same point and having the same duration $T$, as in \fref{Fig1}. Let us first observe that edges of the global convex hull will be of two types: edges joining two points of the same path $\Gamma _{B_i}$ (type $a$) and edges joining two points belonging to two distinct paths $\Gamma _{B_i}$, $\Gamma _{B_j}$ with $i\neq j$ (type $b$), as shown in \fref{FigEdges}.
Therefore, one has to consider not only line segments of the form $[B_i(\tau)B_i(\tau +s)]$ with $i\in \{1,2,\dots,n\}$, $s\in[\Delta t,T]$ and $\tau \in [0,T-s]$, but also line segments of the form $[B_i(\tau_1)B_j(\tau_2)]$ with $i\neq j \in \{1,2,\dots,n\}$, $\tau_1 \in [0,T]$, $\tau_2 \in [0,T]$ and $(\tau_1+\tau_2)\in[\Delta t,2T]$.
\medskip

The reasoning is very similar to that in the $n=1$~-~case:
\begin{enumerate}
\item for type-$a$ edges, we work in the $(i,s,\tau)$-adapted coordinate system, which is simply the $(s,\tau)$-adapted system for the $i^{th}$ motion. The ordinates of the $n$ paths are linear Brownian motions $y_1, y_2,\dots,y_n$ which all start at the same point $y_0 >0$; $y_i$ satisfies the same constraints C1--C3 as $y$ in section~\ref{case1}, and the $(n-1)$ other Brownian motions are simply constrained to remain strictly positive up to time $T$ as shown in \fref{fig5a};
\item for type-$b$ edges, we work in an ``$(i,j,\tau_1,\tau_2)$-adapted'' coordinate system (the $x$-axis of which is defined by the line through $B_i(\tau_1)$ and $B_j(\tau_2)$). $y_1, y_2,\dots,y_n$ start at $y_0 >0$; $y_i$ hits $0$ only once at time $t=\tau_1$ and $y_j$ only once at time $\tau_2$, while the $(n-2)$ other motions remain strictly positive throughout, as shown in \fref{fig5b}.
\end{enumerate}
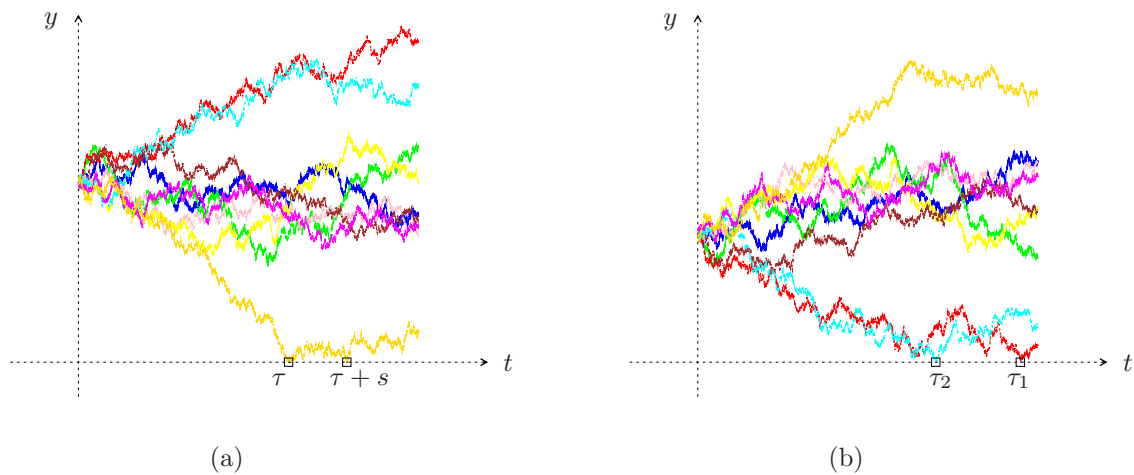
\begin{figure} 
\subfloat[]{ 
\label{fig5a} 
\begin{minipage}[b]{0.45\linewidth} 
\centering \input{Fig5a.tex}
\end{minipage}} 
\hspace{0.9cm}
\subfloat[]{ 
\label{fig5b}
\begin{minipage}[b]{0.45\linewidth} 
\centering \input{Fig5b.tex}
\end{minipage}} 
\caption{The $y$-coordinates of the $n=9$ paths shown on \fref{Fig1} in the coordinate system adapted (a) to the edge of type~$a$ shown on \fref{FigEdges} and (b) to the edge of type~$b$ shown on \fref{FigEdges}. In both cases, all $n$ motions start from the same positive value. In the system adapted to a type-$a$ edge, $(n-1)$ motions are simply constrained to remain positive, while the one motion corresponding to the edge satisfies the same constraints as in section~\ref{case1}, shown in~\fref{fig3b}: it hits $0$ at times $\tau$ and $\tau+s$, and remains positive otherwise. In the system adapted to a type-$b$ edge, the $2$ motions corresponding to the edge hit $0$ just once each (at times $\tau_1$ and $\tau_2$) and remain otherwise positive, while the other $(n-2)$ motions simply remain positive.} 
\label{Fig5}
\end{figure} 

Again, the idea is to compute the probability of the $2$-dimensional configurations by ``counting'' allowed one-dimensional configurations, for fixed $s$ and $\tau$ in the case of type-$a$ edges, and for fixed $\tau_1$ and $\tau_2$ in the case of type-$b$ edges.

Using independence of the $n$ motions, one can express the corresponding probabilities in terms of known propagators: $p$ and $g$ as defined in section~\ref{case1}, along with $q(t;x,y)=p(t;x,y)-p(t;-x,y)$. ($q$ is the Brownian propagator between $x>0$ and $y>0$ in time $t$ under the constraint of not crossing $0$). One obtains:
\begin{enumerate}
\item for type-$a$ edges (there are $n$ paths to choose from):
\begin{eqnarray}
\fl f_n^{(a)}(s,\tau)= 2 \times n\times \frac{\int_0^\infty\,g(\tau ; 0,y_0)\left[\int_0^\infty\, q(T;y_0,v)\,\rmd v\right]^{n-1}\,\rmd y_0 }{\int_0^\infty p(\tau ; z_0,0) \left[\int_{-\infty}^\infty\, p(T;z_0,v')\,\rmd v'\right]^{n-1}\, \rmd z_0}\ \times \nonumber\\
\hspace{-1.5cm} \frac{\int_0^\infty \, g(\kappa-\tau ;0,r)\,g(\tau+s-\kappa ; 0,r)\,\rmd r}{\int_0^\infty \, p(\kappa-\tau ; 0,r')\,p(\tau +s-\kappa ; r',0)\,\rmd r'}\ \times \ 
\frac{\int_0^\infty \, g(T-(\tau +s);0,u)\,\rmd u}{\int_{-\infty}^\infty \, p(T-(\tau +s); 0,u')\, \rmd u'},\label{nDa}
\end{eqnarray}
this is a term of a form very similar to \eref{1Dens}, the only difference being in the first factor, which incorporates the probability that the $(n-1)$ other $y$-motions start from the same point $y_0$ and remain positive throughout their common duration $T$;
\item for type-$b$ edges (there are $n(n-1)$ ordered pairs of paths to choose from):
\begin{eqnarray}
\fl f_n^{(b)}(s,\tau_1,\tau_2)= 2 \times n(n-1)\times \delta(\tau_1+\tau_2-s) \times \nonumber \\ \frac{\int_0^\infty\,g(\tau_1 ; 0,y_0)g(\tau_2 ;0,y_0)\left[\int_0^\infty\, q(T;y_0,v)\,\rmd v\right]^{n-2}\,\rmd y_0 }{\int_0^\infty p(\tau_1 ; z_0,0)p(\tau_2 ; z_0,0) \left[\int_{-\infty}^\infty\, p(T;z_0,v')\,\rmd v'\right]^{n-2}\, \rmd z_0}\ \times \nonumber \\
\frac{\int_0^\infty \, g(T-\tau_1 ;0,u)\,\rmd u}{\int_{-\infty}^\infty \, p(T-\tau_1; 0,u')\, \rmd u'}\times \frac{\int_0^\infty \, g(T-\tau_2 ;0,w)\,\rmd w}{\int_{-\infty}^\infty \, p(T-\tau_2; 0,w')\, \rmd w'},\label{nDb}
\end{eqnarray}
this term is of course specific to $n>1$ ; the $\delta$-function in the first line simply ensures that the ``Brownian-path time'' along the line segment (one can go from end of the line segment to the other following a continuous Brownian path of duration $\tau_1+\tau_2$) is equal to a given $s\in [\Delta t, 2T]$; the following factors correpond to the probability that, having started at some $y_0>0$, $B_i$ (resp. $B_j$) hits $0$ at $\tau _1$ (resp. $\tau_2$) and remains positive before and after, while all other $(n-2)$ $y$-motions start also from $y_0$ and never hit $0$ for their entire duration $T$.
\end{enumerate}

Substituting the expressions for $p$, $q$ and $g$ in equations (\ref{nDa}) and (\ref{nDb}), tidying up the notations and changing to dimensionless variables in the integrals yield:
\begin{eqnarray}
\fl \alpha _n(T,\Delta t) &\equiv \int_{\Delta t}^T\, \int_0^{T-s}\,f_n^{(a)}\,\rmd \tau\,\rmd s &= \int_{\frac{\Delta t}{T}}^{1}\,\int_0^{1-v}\int_0^\infty h_n^{(a)}(u,v,z)\, \rmd u\, \rmd z\, \rmd v,\nonumber\\
& &\label{alpha}\\
\fl \beta _n(T,\Delta t) &\equiv \int_{\Delta t}^T\, \int_0^{T-s}\,f_n^{(b)}\,\rmd \tau\,\rmd s &= \int_{\frac{\Delta t}{T}}^{2}\,\int_{\max(0,v-1)}^{\min (1,v)}\int_0^\infty h_n^{(b)}(u,v,z)\, \rmd u\, \rmd z\, \rmd v,\nonumber\\
& &\label{beta}
\end{eqnarray}
with:
\begin{eqnarray}
&h_n^{(a)}(u,v,z)&= 4n\,\frac{u \exp \left(-\frac{u^2}{z}\right)[\mathrm{erf}(u)]^{n-1}}{\pi v z\left[z(1-v-z)\right]^{\frac{1}{2}}}\label{hna}\\
\fl \rm{and} \nonumber\\
&h_n^{(b)}(u,v,z)&= 4n(n-1)\,\frac{v^{\frac{1}{2}}u^2\exp\left(-\frac{vu^2}{z(v-z)}\right)[\mathrm{erf}(u)]^{n-2}}{\pi z(v-z)\,\left[\pi z(1-z)(v-z)(1-v+z)\right]^{\frac{1}{2}}}\label{hnb}
\end{eqnarray}
where $\mathrm{erf}(u)=\frac{2}{\sqrt{\pi}}\int_0^u\, e^{-x^2}\, \rmd x$ is the error function.
\medskip

Finally, combining equations~(\ref{alpha}) and (\ref{beta}) leads to our main result, an exact formula for the average number of edges (``cut off'' at $\Delta t$) on $\partial \mathcal{C}_n$:
\begin{equation}
\langle V_{n}(T,\Delta t)\rangle  = \alpha _n(T,\Delta t) + \beta _n(T,\Delta t) \label{GF}
\end{equation}
\medskip

A first, simple check of this formula is readily carried out by looking at the case $n=2$, for which the integrals over $u$ are easily performed. We find:
\begin{eqnarray}
\langle V_{2}(T,\Delta t)\rangle &=& \alpha _2(T,\Delta t) + \beta _2(T,\Delta t)\nonumber \\
&=&\frac{4}{\pi} \int_{\frac{\Delta t}{T}}^{1}\,\int_0^{1-v} \left[v^2(z+1)(1-v-z)\right]^{-\frac{1}{2}}\, \rmd z\, \rmd v\nonumber\\
& & + \frac{2}{\pi} \int_{\frac{\Delta t}{T}}^{2}\,\int_{\max(0,v-1)}^{\min (1,v)} \left[v^2(1-z)(1-v+z)\right]^{-\frac{1}{2}}\, \rmd z\, \rmd v \nonumber\\
&=& 2\, \ln \left( \frac{2T}{\Delta t}\right),
\end{eqnarray}
as expected indeed, since the case $n=2$ is essentially the same, as far as the convex hull is concerned, as that of a single path of double duration.

\section{A new derivation of the average-perimeter formula}
\label{cper}
Another confirmation of equation~(\ref{GF}) can be obtained by using the underlying probability distribution to retrieve the exact formula for the average perimeter $\left\langle L_n(T)\right\rangle$ of $\partial \mathcal{C}_n$ which was derived using an entirely different method in \cite{RFSMAC}:
\begin{equation}
\left\langle L_n(T)\right\rangle = 4n \sqrt{2\pi T}\int_0^\infty u\ e^{-u^2}\ [\mathrm{erf}(u)]^{n-1}\,\rmd u. \label{Ln}
\end{equation}

Indeed, multiplying the probability that a given line segment belongs to $\partial \mathcal{C}_n$ by the average Euclidean length of this line segment before summing over all possible line segments should allow one to take the limit $\Delta t = 0$ and re-obtain \eref{Ln}.

Let us therefore first compute the expected Euclidean length $l(s)$ of the line segment $[B(\tau)B(\tau+s)]$ (by the Markovian nature of Brownian motion, this average length is independent of $\tau$). This should not be done in the $(s,\tau)$-adapted frame, as this frame itself varies from one realization to another. Instead, one should use some fixed frame, \textit{eg} the canonical basis of $\mathbb{R}^2$, in which the question now simply amounts to computing the average norm of a bivariate normal distribution with zero-correlation. This is a standard exercise and one finds that:
\begin{equation}
l(s)= \sqrt{\frac{\pi s}{2}}.
\end{equation}
Note that for line segments of type~$b$ which are such that $\tau_1+\tau_2=s$, the average length also depends only on $s$ and has the same expression.

We shall now compute the expected perimeter length of  $\partial \mathcal{C}_n$ according to the probabilities established in section~\ref{casegen}:
\begin{eqnarray}
\fl \left\langle L_n(T)\right\rangle =\nonumber \\ \fl \int_0^\infty\left(\int_{\frac{\Delta t}{T}}^{1}\int_0^{1-v}\,l(vT)\,h_n^{(a)}(u,v,z)\, \rmd z\, \rmd v + \int_{\frac{\Delta t}{T}}^{2}\int_{\max(0,v-1)}^{\min (1,v)}\,l(vT)\,h_n^{(b)}(u,v,z)\, \rmd z\, \rmd v\right)\, \rmd u \nonumber\\
\label{LnInt}
\end{eqnarray}
Once multiplied by $l(vT)$, the integrands in \eref{LnInt} become integrable from $v=0$, meaning that we can set $\Delta t=0$ as expected. A few changes of variables then lead to a form where the integrals on $v$ and $z$ can be performed (see also \cite{GR}). One obtains:
\begin{eqnarray}
\fl \left\langle L_n(T)\right\rangle = 4n\sqrt{2\pi T}\times\nonumber \\
\nonumber \\
\hspace{-2cm}\left[\frac{\sqrt{\pi}}{2}\int_0^{\infty}\left([\mathrm{erf}(u)]^{n-1}-[\mathrm{erf}(u)]^{n}\right)\, \rmd u + (n-1)\int_0^{\infty}u\left(1-\mathrm{erf}(u)\right) e^{-u^2}[\mathrm{erf}(u)]^{n-2}\, \rmd u\right].\nonumber\\
\label{Lnab}
\end{eqnarray}
A simple integration by parts of the second term in \eref{Lnab} then gives:
\begin{equation}
\left\langle L_n(T)\right\rangle = 4n \sqrt{2\pi T}\int_0^\infty u\ e^{-u^2}\ [\mathrm{erf}(u)]^{n-1}\,\rmd u, \label{Ln2}
\end{equation}
exactly as in \eref{Ln}, thus confirming our main result, \eref{GF}.

\section*{Conclusion}

More work is needed to give \eref{GF} a more explicit form, but to the best of our knowledge this is the first exact formula for the number of edges on the boundary of the convex hull of planar Brownian paths. Progress on the treatment of the triple integrals appearing in \eref{alpha} and \eref{beta}, including numerical evaluation and asymptotical approximations will allow comparison with results of computer simulations (which we have not included here).

Equivalent formulae for other types of stochastic processes, be they continuous-time or dicrete-time random walks in the plane, might be obtained by transposing some of the steps followed here. Further mathematical formalization of the probabilistic objects defined and used in this paper will be pursued and should lead to insights regarding potential extensions.

Future work will also include looking at how the ideas introduced here, in particular the Brownian-path time ``cut off'', translate into the path-integral formalism which has proven quite powerful in contexts of constrained Brownian motions \cite{ACSM1,MKRF}. Such translation could be of interest especially in relationship with the study of Brownian cone points \cite{Ev,Shi2,Bu,LeG,ACJDProp}.

\ack
The author wishes to thank Kirone Mallick for his encouragement and for his comments on this manuscript.

\section*{References}
\bibliographystyle{unsrt}
\bibliography{BMCH}

\end{document}

%% file: Fig1.tex
\begingroup
  \makeatletter
  \providecommand\color[2][]{%
    \GenericError{(gnuplot) \space\space\space\@spaces}{%
      Package color not loaded in conjunction with
      terminal option `colourtext'%
    }{See the gnuplot documentation for explanation.%
    }{Either use 'blacktext' in gnuplot or load the package
      color.sty in LaTeX.}%
    \renewcommand\color[2][]{}%
  }%
  \providecommand\includegraphics[2][]{%
    \GenericError{(gnuplot) \space\space\space\@spaces}{%
      Package graphicx or graphics not loaded%
    }{See the gnuplot documentation for explanation.%
    }{The gnuplot epslatex terminal needs graphicx.sty or graphics.sty.}%
    \renewcommand\includegraphics[2][]{}%
  }%
  \providecommand\rotatebox[2]{#2}%
  \@ifundefined{ifGPcolor}{%
    \newif\ifGPcolor
    \GPcolortrue
  }{}%
  \@ifundefined{ifGPblacktext}{%
    \newif\ifGPblacktext
    \GPblacktexttrue
  }{}%
  \let\gplgaddtomacro\g@addto@macro
  \gdef\gplbacktext{}%
  \gdef\gplfronttext{}%
  \makeatother
  \ifGPblacktext
    \def\colorrgb#1{}%
    \def\colorgray#1{}%
  \else
    \ifGPcolor
      \def\colorrgb#1{\color[rgb]{#1}}%
      \def\colorgray#1{\color[gray]{#1}}%
      \expandafter\def\csname LTw\endcsname{\color{white}}%
      \expandafter\def\csname LTb\endcsname{\color{black}}%
      \expandafter\def\csname LTa\endcsname{\color{black}}%
      \expandafter\def\csname LT0\endcsname{\color[rgb]{1,0,0}}%
      \expandafter\def\csname LT1\endcsname{\color[rgb]{0,1,0}}%
      \expandafter\def\csname LT2\endcsname{\color[rgb]{0,0,1}}%
      \expandafter\def\csname LT3\endcsname{\color[rgb]{1,0,1}}%
      \expandafter\def\csname LT4\endcsname{\color[rgb]{0,1,1}}%
      \expandafter\def\csname LT5\endcsname{\color[rgb]{1,1,0}}%
      \expandafter\def\csname LT6\endcsname{\color[rgb]{0,0,0}}%
      \expandafter\def\csname LT7\endcsname{\color[rgb]{1,0.3,0}}%
      \expandafter\def\csname LT8\endcsname{\color[rgb]{0.5,0.5,0.5}}%
    \else
      \def\colorrgb#1{\color{black}}%
      \def\colorgray#1{\color[gray]{#1}}%
      \expandafter\def\csname LTw\endcsname{\color{white}}%
      \expandafter\def\csname LTb\endcsname{\color{black}}%
      \expandafter\def\csname LTa\endcsname{\color{black}}%
      \expandafter\def\csname LT0\endcsname{\color{black}}%
      \expandafter\def\csname LT1\endcsname{\color{black}}%
      \expandafter\def\csname LT2\endcsname{\color{black}}%
      \expandafter\def\csname LT3\endcsname{\color{black}}%
      \expandafter\def\csname LT4\endcsname{\color{black}}%
      \expandafter\def\csname LT5\endcsname{\color{black}}%
      \expandafter\def\csname LT6\endcsname{\color{black}}%
      \expandafter\def\csname LT7\endcsname{\color{black}}%
      \expandafter\def\csname LT8\endcsname{\color{black}}%
    \fi
  \fi
  \setlength{\unitlength}{0.0500bp}%
  \begin{picture}(5668.00,4250.00)%
    \gplgaddtomacro\gplbacktext{%
    }%
    \gplgaddtomacro\gplfronttext{%
    }%
    \gplbacktext
    \put(0,0){\includegraphics{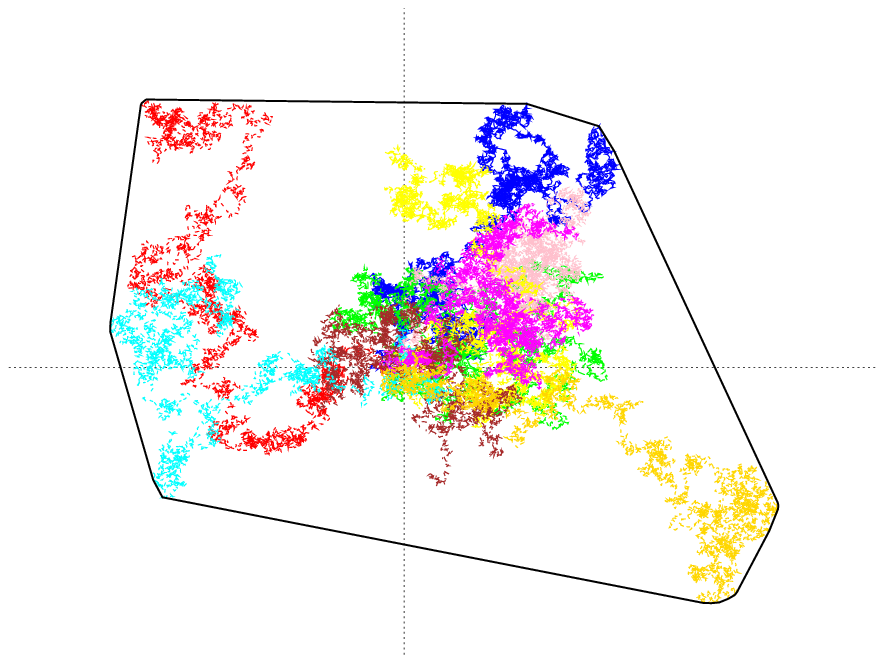}}%
    \gplfronttext
  \end{picture}%
\endgroup

%% file: Fig2a.tex
\begingroup
  \makeatletter
  \providecommand\color[2][]{%
    \GenericError{(gnuplot) \space\space\space\@spaces}{%
      Package color not loaded in conjunction with
      terminal option `colourtext'%
    }{See the gnuplot documentation for explanation.%
    }{Either use 'blacktext' in gnuplot or load the package
      color.sty in LaTeX.}%
    \renewcommand\color[2][]{}%
  }%
  \providecommand\includegraphics[2][]{%
    \GenericError{(gnuplot) \space\space\space\@spaces}{%
      Package graphicx or graphics not loaded%
    }{See the gnuplot documentation for explanation.%
    }{The gnuplot epslatex terminal needs graphicx.sty or graphics.sty.}%
    \renewcommand\includegraphics[2][]{}%
  }%
  \providecommand\rotatebox[2]{#2}%
  \@ifundefined{ifGPcolor}{%
    \newif\ifGPcolor
    \GPcolortrue
  }{}%
  \@ifundefined{ifGPblacktext}{%
    \newif\ifGPblacktext
    \GPblacktexttrue
  }{}%
  \let\gplgaddtomacro\g@addto@macro
  \gdef\gplbacktext{}%
  \gdef\gplfronttext{}%
  \makeatother
  \ifGPblacktext
    \def\colorrgb#1{}%
    \def\colorgray#1{}%
  \else
    \ifGPcolor
      \def\colorrgb#1{\color[rgb]{#1}}%
      \def\colorgray#1{\color[gray]{#1}}%
      \expandafter\def\csname LTw\endcsname{\color{white}}%
      \expandafter\def\csname LTb\endcsname{\color{black}}%
      \expandafter\def\csname LTa\endcsname{\color{black}}%
      \expandafter\def\csname LT0\endcsname{\color[rgb]{1,0,0}}%
      \expandafter\def\csname LT1\endcsname{\color[rgb]{0,1,0}}%
      \expandafter\def\csname LT2\endcsname{\color[rgb]{0,0,1}}%
      \expandafter\def\csname LT3\endcsname{\color[rgb]{1,0,1}}%
      \expandafter\def\csname LT4\endcsname{\color[rgb]{0,1,1}}%
      \expandafter\def\csname LT5\endcsname{\color[rgb]{1,1,0}}%
      \expandafter\def\csname LT6\endcsname{\color[rgb]{0,0,0}}%
      \expandafter\def\csname LT7\endcsname{\color[rgb]{1,0.3,0}}%
      \expandafter\def\csname LT8\endcsname{\color[rgb]{0.5,0.5,0.5}}%
    \else
      \def\colorrgb#1{\color{black}}%
      \def\colorgray#1{\color[gray]{#1}}%
      \expandafter\def\csname LTw\endcsname{\color{white}}%
      \expandafter\def\csname LTb\endcsname{\color{black}}%
      \expandafter\def\csname LTa\endcsname{\color{black}}%
      \expandafter\def\csname LT0\endcsname{\color{black}}%
      \expandafter\def\csname LT1\endcsname{\color{black}}%
      \expandafter\def\csname LT2\endcsname{\color{black}}%
      \expandafter\def\csname LT3\endcsname{\color{black}}%
      \expandafter\def\csname LT4\endcsname{\color{black}}%
      \expandafter\def\csname LT5\endcsname{\color{black}}%
      \expandafter\def\csname LT6\endcsname{\color{black}}%
      \expandafter\def\csname LT7\endcsname{\color{black}}%
      \expandafter\def\csname LT8\endcsname{\color{black}}%
    \fi
  \fi
  \setlength{\unitlength}{0.0500bp}%
  \begin{picture}(3118.00,3118.00)%
    \gplgaddtomacro\gplbacktext{%
      \csname LTb\endcsname%
      \put(163,852){\makebox(0,0)[l]{\strut{}\footnotesize $O$}}%
      \put(879,571){\makebox(0,0)[l]{\strut{}\footnotesize $B(\tau)$}}%
      \put(1559,1274){\makebox(0,0)[l]{\strut{}\footnotesize $B(\tau +s)$}}%
    }%
    \gplgaddtomacro\gplfronttext{%
    }%
    \gplbacktext
    \put(0,0){\includegraphics{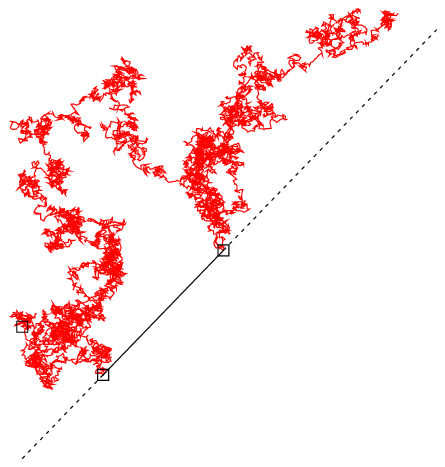}}%
    \gplfronttext
  \end{picture}%
\endgroup

%% file: Fig2b.tex
\begingroup
  \makeatletter
  \providecommand\color[2][]{%
    \GenericError{(gnuplot) \space\space\space\@spaces}{%
      Package color not loaded in conjunction with
      terminal option `colourtext'%
    }{See the gnuplot documentation for explanation.%
    }{Either use 'blacktext' in gnuplot or load the package
      color.sty in LaTeX.}%
    \renewcommand\color[2][]{}%
  }%
  \providecommand\includegraphics[2][]{%
    \GenericError{(gnuplot) \space\space\space\@spaces}{%
      Package graphicx or graphics not loaded%
    }{See the gnuplot documentation for explanation.%
    }{The gnuplot epslatex terminal needs graphicx.sty or graphics.sty.}%
    \renewcommand\includegraphics[2][]{}%
  }%
  \providecommand\rotatebox[2]{#2}%
  \@ifundefined{ifGPcolor}{%
    \newif\ifGPcolor
    \GPcolortrue
  }{}%
  \@ifundefined{ifGPblacktext}{%
    \newif\ifGPblacktext
    \GPblacktexttrue
  }{}%
  \let\gplgaddtomacro\g@addto@macro
  \gdef\gplbacktext{}%
  \gdef\gplfronttext{}%
  \makeatother
  \ifGPblacktext
    \def\colorrgb#1{}%
    \def\colorgray#1{}%
  \else
    \ifGPcolor
      \def\colorrgb#1{\color[rgb]{#1}}%
      \def\colorgray#1{\color[gray]{#1}}%
      \expandafter\def\csname LTw\endcsname{\color{white}}%
      \expandafter\def\csname LTb\endcsname{\color{black}}%
      \expandafter\def\csname LTa\endcsname{\color{black}}%
      \expandafter\def\csname LT0\endcsname{\color[rgb]{1,0,0}}%
      \expandafter\def\csname LT1\endcsname{\color[rgb]{0,1,0}}%
      \expandafter\def\csname LT2\endcsname{\color[rgb]{0,0,1}}%
      \expandafter\def\csname LT3\endcsname{\color[rgb]{1,0,1}}%
      \expandafter\def\csname LT4\endcsname{\color[rgb]{0,1,1}}%
      \expandafter\def\csname LT5\endcsname{\color[rgb]{1,1,0}}%
      \expandafter\def\csname LT6\endcsname{\color[rgb]{0,0,0}}%
      \expandafter\def\csname LT7\endcsname{\color[rgb]{1,0.3,0}}%
      \expandafter\def\csname LT8\endcsname{\color[rgb]{0.5,0.5,0.5}}%
    \else
      \def\colorrgb#1{\color{black}}%
      \def\colorgray#1{\color[gray]{#1}}%
      \expandafter\def\csname LTw\endcsname{\color{white}}%
      \expandafter\def\csname LTb\endcsname{\color{black}}%
      \expandafter\def\csname LTa\endcsname{\color{black}}%
      \expandafter\def\csname LT0\endcsname{\color{black}}%
      \expandafter\def\csname LT1\endcsname{\color{black}}%
      \expandafter\def\csname LT2\endcsname{\color{black}}%
      \expandafter\def\csname LT3\endcsname{\color{black}}%
      \expandafter\def\csname LT4\endcsname{\color{black}}%
      \expandafter\def\csname LT5\endcsname{\color{black}}%
      \expandafter\def\csname LT6\endcsname{\color{black}}%
      \expandafter\def\csname LT7\endcsname{\color{black}}%
      \expandafter\def\csname LT8\endcsname{\color{black}}%
    \fi
  \fi
  \setlength{\unitlength}{0.0500bp}%
  \begin{picture}(4250.00,2834.00)%
    \gplgaddtomacro\gplbacktext{%
      \csname LTb\endcsname%
      \put(384,353){\makebox(0,0)[l]{\strut{}\footnotesize $O$}}%
      \put(1198,861){\makebox(0,0)[l]{\strut{}\footnotesize $B(\tau)$}}%
      \put(2473,1652){\makebox(0,0)[l]{\strut{}\footnotesize $B(\tau +s)$}}%
    }%
    \gplgaddtomacro\gplfronttext{%
    }%
    \gplbacktext
    \put(0,0){\includegraphics{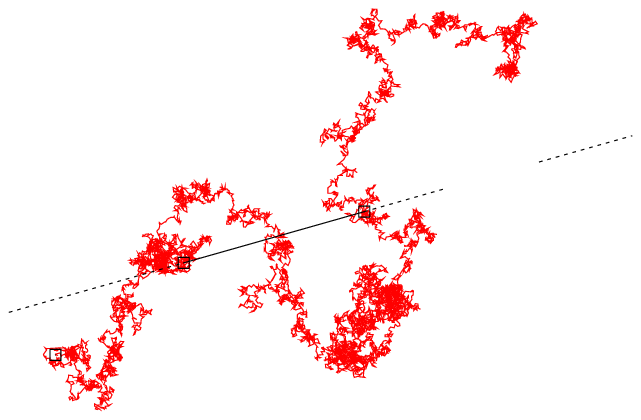}}%
    \gplfronttext
  \end{picture}%
\endgroup

%% file: Fig3a.tex
\begingroup
  \makeatletter
  \providecommand\color[2][]{%
    \GenericError{(gnuplot) \space\space\space\@spaces}{%
      Package color not loaded in conjunction with
      terminal option `colourtext'%
    }{See the gnuplot documentation for explanation.%
    }{Either use 'blacktext' in gnuplot or load the package
      color.sty in LaTeX.}%
    \renewcommand\color[2][]{}%
  }%
  \providecommand\includegraphics[2][]{%
    \GenericError{(gnuplot) \space\space\space\@spaces}{%
      Package graphicx or graphics not loaded%
    }{See the gnuplot documentation for explanation.%
    }{The gnuplot epslatex terminal needs graphicx.sty or graphics.sty.}%
    \renewcommand\includegraphics[2][]{}%
  }%
  \providecommand\rotatebox[2]{#2}%
  \@ifundefined{ifGPcolor}{%
    \newif\ifGPcolor
    \GPcolortrue
  }{}%
  \@ifundefined{ifGPblacktext}{%
    \newif\ifGPblacktext
    \GPblacktexttrue
  }{}%
  \let\gplgaddtomacro\g@addto@macro
  \gdef\gplbacktext{}%
  \gdef\gplfronttext{}%
  \makeatother
  \ifGPblacktext
    \def\colorrgb#1{}%
    \def\colorgray#1{}%
  \else
    \ifGPcolor
      \def\colorrgb#1{\color[rgb]{#1}}%
      \def\colorgray#1{\color[gray]{#1}}%
      \expandafter\def\csname LTw\endcsname{\color{white}}%
      \expandafter\def\csname LTb\endcsname{\color{black}}%
      \expandafter\def\csname LTa\endcsname{\color{black}}%
      \expandafter\def\csname LT0\endcsname{\color[rgb]{1,0,0}}%
      \expandafter\def\csname LT1\endcsname{\color[rgb]{0,1,0}}%
      \expandafter\def\csname LT2\endcsname{\color[rgb]{0,0,1}}%
      \expandafter\def\csname LT3\endcsname{\color[rgb]{1,0,1}}%
      \expandafter\def\csname LT4\endcsname{\color[rgb]{0,1,1}}%
      \expandafter\def\csname LT5\endcsname{\color[rgb]{1,1,0}}%
      \expandafter\def\csname LT6\endcsname{\color[rgb]{0,0,0}}%
      \expandafter\def\csname LT7\endcsname{\color[rgb]{1,0.3,0}}%
      \expandafter\def\csname LT8\endcsname{\color[rgb]{0.5,0.5,0.5}}%
    \else
      \def\colorrgb#1{\color{black}}%
      \def\colorgray#1{\color[gray]{#1}}%
      \expandafter\def\csname LTw\endcsname{\color{white}}%
      \expandafter\def\csname LTb\endcsname{\color{black}}%
      \expandafter\def\csname LTa\endcsname{\color{black}}%
      \expandafter\def\csname LT0\endcsname{\color{black}}%
      \expandafter\def\csname LT1\endcsname{\color{black}}%
      \expandafter\def\csname LT2\endcsname{\color{black}}%
      \expandafter\def\csname LT3\endcsname{\color{black}}%
      \expandafter\def\csname LT4\endcsname{\color{black}}%
      \expandafter\def\csname LT5\endcsname{\color{black}}%
      \expandafter\def\csname LT6\endcsname{\color{black}}%
      \expandafter\def\csname LT7\endcsname{\color{black}}%
      \expandafter\def\csname LT8\endcsname{\color{black}}%
    \fi
  \fi
  \setlength{\unitlength}{0.0500bp}%
  \begin{picture}(4250.00,2834.00)%
    \gplgaddtomacro\gplbacktext{%
      \csname LTb\endcsname%
      \put(4040,671){\makebox(0,0)[l]{\strut{}\footnotesize $x$}}%
      \put(569,2570){\makebox(0,0)[l]{\strut{}\footnotesize $y$}}%
      \put(330,1363){\makebox(0,0)[l]{\strut{}\footnotesize $B(0)$}}%
      \put(842,467){\makebox(0,0)[l]{\strut{}\footnotesize $B(\tau)$}}%
      \put(1886,467){\makebox(0,0)[l]{\strut{}\footnotesize $B(\tau +s)$}}%
    }%
    \gplgaddtomacro\gplfronttext{%
    }%
    \gplbacktext
    \put(0,0){\includegraphics{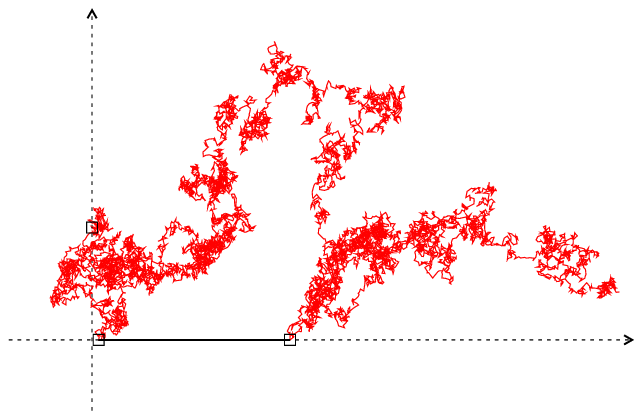}}%
    \gplfronttext
  \end{picture}%
\endgroup

%% file: Fig3b.tex
\begingroup
  \makeatletter
  \providecommand\color[2][]{%
    \GenericError{(gnuplot) \space\space\space\@spaces}{%
      Package color not loaded in conjunction with
      terminal option `colourtext'%
    }{See the gnuplot documentation for explanation.%
    }{Either use 'blacktext' in gnuplot or load the package
      color.sty in LaTeX.}%
    \renewcommand\color[2][]{}%
  }%
  \providecommand\includegraphics[2][]{%
    \GenericError{(gnuplot) \space\space\space\@spaces}{%
      Package graphicx or graphics not loaded%
    }{See the gnuplot documentation for explanation.%
    }{The gnuplot epslatex terminal needs graphicx.sty or graphics.sty.}%
    \renewcommand\includegraphics[2][]{}%
  }%
  \providecommand\rotatebox[2]{#2}%
  \@ifundefined{ifGPcolor}{%
    \newif\ifGPcolor
    \GPcolortrue
  }{}%
  \@ifundefined{ifGPblacktext}{%
    \newif\ifGPblacktext
    \GPblacktexttrue
  }{}%
  \let\gplgaddtomacro\g@addto@macro
  \gdef\gplbacktext{}%
  \gdef\gplfronttext{}%
  \makeatother
  \ifGPblacktext
    \def\colorrgb#1{}%
    \def\colorgray#1{}%
  \else
    \ifGPcolor
      \def\colorrgb#1{\color[rgb]{#1}}%
      \def\colorgray#1{\color[gray]{#1}}%
      \expandafter\def\csname LTw\endcsname{\color{white}}%
      \expandafter\def\csname LTb\endcsname{\color{black}}%
      \expandafter\def\csname LTa\endcsname{\color{black}}%
      \expandafter\def\csname LT0\endcsname{\color[rgb]{1,0,0}}%
      \expandafter\def\csname LT1\endcsname{\color[rgb]{0,1,0}}%
      \expandafter\def\csname LT2\endcsname{\color[rgb]{0,0,1}}%
      \expandafter\def\csname LT3\endcsname{\color[rgb]{1,0,1}}%
      \expandafter\def\csname LT4\endcsname{\color[rgb]{0,1,1}}%
      \expandafter\def\csname LT5\endcsname{\color[rgb]{1,1,0}}%
      \expandafter\def\csname LT6\endcsname{\color[rgb]{0,0,0}}%
      \expandafter\def\csname LT7\endcsname{\color[rgb]{1,0.3,0}}%
      \expandafter\def\csname LT8\endcsname{\color[rgb]{0.5,0.5,0.5}}%
    \else
      \def\colorrgb#1{\color{black}}%
      \def\colorgray#1{\color[gray]{#1}}%
      \expandafter\def\csname LTw\endcsname{\color{white}}%
      \expandafter\def\csname LTb\endcsname{\color{black}}%
      \expandafter\def\csname LTa\endcsname{\color{black}}%
      \expandafter\def\csname LT0\endcsname{\color{black}}%
      \expandafter\def\csname LT1\endcsname{\color{black}}%
      \expandafter\def\csname LT2\endcsname{\color{black}}%
      \expandafter\def\csname LT3\endcsname{\color{black}}%
      \expandafter\def\csname LT4\endcsname{\color{black}}%
      \expandafter\def\csname LT5\endcsname{\color{black}}%
      \expandafter\def\csname LT6\endcsname{\color{black}}%
      \expandafter\def\csname LT7\endcsname{\color{black}}%
      \expandafter\def\csname LT8\endcsname{\color{black}}%
    \fi
  \fi
  \setlength{\unitlength}{0.0500bp}%
  \begin{picture}(4250.00,2834.00)%
    \gplgaddtomacro\gplbacktext{%
      \csname LTb\endcsname%
      \put(4040,671){\makebox(0,0)[l]{\strut{}\footnotesize $t$}}%
      \put(390,2570){\makebox(0,0)[l]{\strut{}\footnotesize $y$}}%
      \put(330,1315){\makebox(0,0)[l]{\strut{}\footnotesize $y_0$}}%
      \put(1018,495){\makebox(0,0)[l]{\strut{}\footnotesize $\tau$}}%
      \put(2783,495){\makebox(0,0)[l]{\strut{}\footnotesize $\tau+s$}}%
    }%
    \gplgaddtomacro\gplfronttext{%
    }%
    \gplbacktext
    \put(0,0){\includegraphics{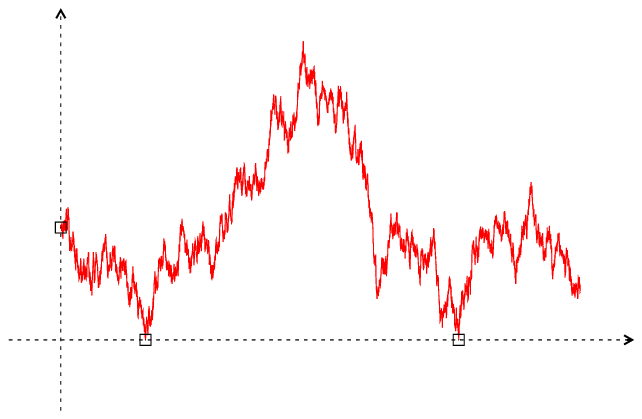}}%
    \gplfronttext
  \end{picture}%
\endgroup

%% file: Fig4.tex
\begingroup
  \makeatletter
  \providecommand\color[2][]{%
    \GenericError{(gnuplot) \space\space\space\@spaces}{%
      Package color not loaded in conjunction with
      terminal option `colourtext'%
    }{See the gnuplot documentation for explanation.%
    }{Either use 'blacktext' in gnuplot or load the package
      color.sty in LaTeX.}%
    \renewcommand\color[2][]{}%
  }%
  \providecommand\includegraphics[2][]{%
    \GenericError{(gnuplot) \space\space\space\@spaces}{%
      Package graphicx or graphics not loaded%
    }{See the gnuplot documentation for explanation.%
    }{The gnuplot epslatex terminal needs graphicx.sty or graphics.sty.}%
    \renewcommand\includegraphics[2][]{}%
  }%
  \providecommand\rotatebox[2]{#2}%
  \@ifundefined{ifGPcolor}{%
    \newif\ifGPcolor
    \GPcolortrue
  }{}%
  \@ifundefined{ifGPblacktext}{%
    \newif\ifGPblacktext
    \GPblacktexttrue
  }{}%
  \let\gplgaddtomacro\g@addto@macro
  \gdef\gplbacktext{}%
  \gdef\gplfronttext{}%
  \makeatother
  \ifGPblacktext
    \def\colorrgb#1{}%
    \def\colorgray#1{}%
  \else
    \ifGPcolor
      \def\colorrgb#1{\color[rgb]{#1}}%
      \def\colorgray#1{\color[gray]{#1}}%
      \expandafter\def\csname LTw\endcsname{\color{white}}%
      \expandafter\def\csname LTb\endcsname{\color{black}}%
      \expandafter\def\csname LTa\endcsname{\color{black}}%
      \expandafter\def\csname LT0\endcsname{\color[rgb]{1,0,0}}%
      \expandafter\def\csname LT1\endcsname{\color[rgb]{0,1,0}}%
      \expandafter\def\csname LT2\endcsname{\color[rgb]{0,0,1}}%
      \expandafter\def\csname LT3\endcsname{\color[rgb]{1,0,1}}%
      \expandafter\def\csname LT4\endcsname{\color[rgb]{0,1,1}}%
      \expandafter\def\csname LT5\endcsname{\color[rgb]{1,1,0}}%
      \expandafter\def\csname LT6\endcsname{\color[rgb]{0,0,0}}%
      \expandafter\def\csname LT7\endcsname{\color[rgb]{1,0.3,0}}%
      \expandafter\def\csname LT8\endcsname{\color[rgb]{0.5,0.5,0.5}}%
    \else
      \def\colorrgb#1{\color{black}}%
      \def\colorgray#1{\color[gray]{#1}}%
      \expandafter\def\csname LTw\endcsname{\color{white}}%
      \expandafter\def\csname LTb\endcsname{\color{black}}%
      \expandafter\def\csname LTa\endcsname{\color{black}}%
      \expandafter\def\csname LT0\endcsname{\color{black}}%
      \expandafter\def\csname LT1\endcsname{\color{black}}%
      \expandafter\def\csname LT2\endcsname{\color{black}}%
      \expandafter\def\csname LT3\endcsname{\color{black}}%
      \expandafter\def\csname LT4\endcsname{\color{black}}%
      \expandafter\def\csname LT5\endcsname{\color{black}}%
      \expandafter\def\csname LT6\endcsname{\color{black}}%
      \expandafter\def\csname LT7\endcsname{\color{black}}%
      \expandafter\def\csname LT8\endcsname{\color{black}}%
    \fi
  \fi
  \setlength{\unitlength}{0.0500bp}%
  \begin{picture}(6802.00,5102.00)%
    \gplgaddtomacro\gplbacktext{%
      \csname LTb\endcsname%
      \put(1047,4147){\makebox(0,0)[l]{\strut{}\footnotesize $B_3(\tau _1)$}}%
      \put(842,2612){\makebox(0,0)[l]{\strut{}\footnotesize $B_7(\tau _2)$}}%
      \put(23,3364){\makebox(0,0)[l]{\strut{}\footnotesize Edge of type $b$}}%
      \put(5602,569){\makebox(0,0)[l]{\strut{}\footnotesize $B_9(\tau +s)$}}%
      \put(5858,1280){\makebox(0,0)[l]{\strut{}\footnotesize $B_9(\tau)$}}%
      \put(6267,925){\makebox(0,0)[l]{\strut{}\footnotesize Edge of type $a$}}%
    }%
    \gplgaddtomacro\gplfronttext{%
    }%
    \gplbacktext
    \put(0,0){\includegraphics{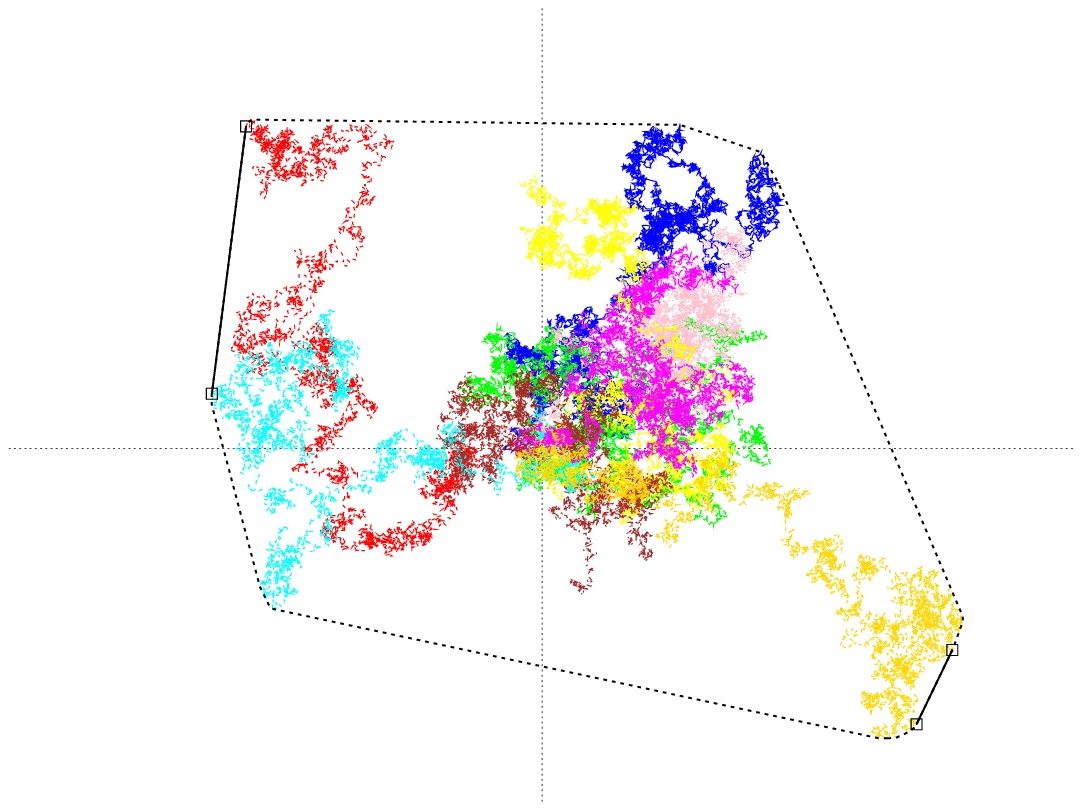}}%
    \gplfronttext
  \end{picture}%
\endgroup

%% file: Fig5a.tex
\begingroup
  \makeatletter
  \providecommand\color[2][]{%
    \GenericError{(gnuplot) \space\space\space\@spaces}{%
      Package color not loaded in conjunction with
      terminal option `colourtext'%
    }{See the gnuplot documentation for explanation.%
    }{Either use 'blacktext' in gnuplot or load the package
      color.sty in LaTeX.}%
    \renewcommand\color[2][]{}%
  }%
  \providecommand\includegraphics[2][]{%
    \GenericError{(gnuplot) \space\space\space\@spaces}{%
      Package graphicx or graphics not loaded%
    }{See the gnuplot documentation for explanation.%
    }{The gnuplot epslatex terminal needs graphicx.sty or graphics.sty.}%
    \renewcommand\includegraphics[2][]{}%
  }%
  \providecommand\rotatebox[2]{#2}%
  \@ifundefined{ifGPcolor}{%
    \newif\ifGPcolor
    \GPcolortrue
  }{}%
  \@ifundefined{ifGPblacktext}{%
    \newif\ifGPblacktext
    \GPblacktexttrue
  }{}%
  \let\gplgaddtomacro\g@addto@macro
  \gdef\gplbacktext{}%
  \gdef\gplfronttext{}%
  \makeatother
  \ifGPblacktext
    \def\colorrgb#1{}%
    \def\colorgray#1{}%
  \else
    \ifGPcolor
      \def\colorrgb#1{\color[rgb]{#1}}%
      \def\colorgray#1{\color[gray]{#1}}%
      \expandafter\def\csname LTw\endcsname{\color{white}}%
      \expandafter\def\csname LTb\endcsname{\color{black}}%
      \expandafter\def\csname LTa\endcsname{\color{black}}%
      \expandafter\def\csname LT0\endcsname{\color[rgb]{1,0,0}}%
      \expandafter\def\csname LT1\endcsname{\color[rgb]{0,1,0}}%
      \expandafter\def\csname LT2\endcsname{\color[rgb]{0,0,1}}%
      \expandafter\def\csname LT3\endcsname{\color[rgb]{1,0,1}}%
      \expandafter\def\csname LT4\endcsname{\color[rgb]{0,1,1}}%
      \expandafter\def\csname LT5\endcsname{\color[rgb]{1,1,0}}%
      \expandafter\def\csname LT6\endcsname{\color[rgb]{0,0,0}}%
      \expandafter\def\csname LT7\endcsname{\color[rgb]{1,0.3,0}}%
      \expandafter\def\csname LT8\endcsname{\color[rgb]{0.5,0.5,0.5}}%
    \else
      \def\colorrgb#1{\color{black}}%
      \def\colorgray#1{\color[gray]{#1}}%
      \expandafter\def\csname LTw\endcsname{\color{white}}%
      \expandafter\def\csname LTb\endcsname{\color{black}}%
      \expandafter\def\csname LTa\endcsname{\color{black}}%
      \expandafter\def\csname LT0\endcsname{\color{black}}%
      \expandafter\def\csname LT1\endcsname{\color{black}}%
      \expandafter\def\csname LT2\endcsname{\color{black}}%
      \expandafter\def\csname LT3\endcsname{\color{black}}%
      \expandafter\def\csname LT4\endcsname{\color{black}}%
      \expandafter\def\csname LT5\endcsname{\color{black}}%
      \expandafter\def\csname LT6\endcsname{\color{black}}%
      \expandafter\def\csname LT7\endcsname{\color{black}}%
      \expandafter\def\csname LT8\endcsname{\color{black}}%
    \fi
  \fi
  \setlength{\unitlength}{0.0500bp}%
  \begin{picture}(4250.00,3400.00)%
    \gplgaddtomacro\gplbacktext{%
      \csname LTb\endcsname%
      \put(4048,525){\makebox(0,0)[l]{\strut{}\footnotesize $t$}}%
      \put(586,3136){\makebox(0,0)[l]{\strut{}\footnotesize $y$}}%
      \put(2304,421){\makebox(0,0)[l]{\strut{}\footnotesize $\tau$}}%
      \put(2740,421){\makebox(0,0)[l]{\strut{}\footnotesize $\tau+s$}}%
    }%
    \gplgaddtomacro\gplfronttext{%
    }%
    \gplbacktext
    \put(0,0){\includegraphics{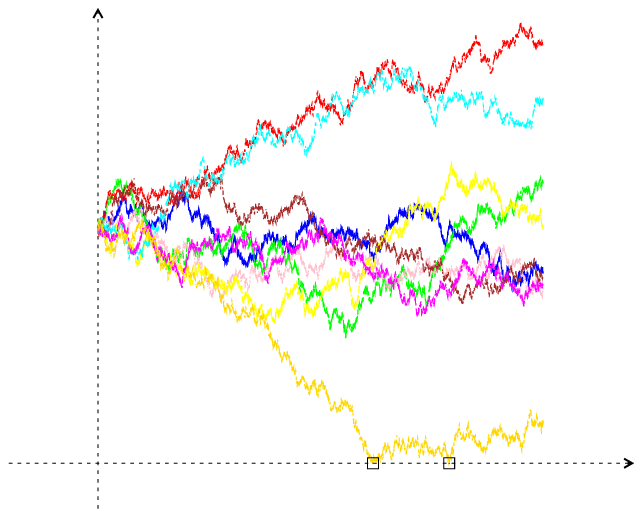}}%
    \gplfronttext
  \end{picture}%
\endgroup

%% file: Fig5b.tex
\begingroup
  \makeatletter
  \providecommand\color[2][]{%
    \GenericError{(gnuplot) \space\space\space\@spaces}{%
      Package color not loaded in conjunction with
      terminal option `colourtext'%
    }{See the gnuplot documentation for explanation.%
    }{Either use 'blacktext' in gnuplot or load the package
      color.sty in LaTeX.}%
    \renewcommand\color[2][]{}%
  }%
  \providecommand\includegraphics[2][]{%
    \GenericError{(gnuplot) \space\space\space\@spaces}{%
      Package graphicx or graphics not loaded%
    }{See the gnuplot documentation for explanation.%
    }{The gnuplot epslatex terminal needs graphicx.sty or graphics.sty.}%
    \renewcommand\includegraphics[2][]{}%
  }%
  \providecommand\rotatebox[2]{#2}%
  \@ifundefined{ifGPcolor}{%
    \newif\ifGPcolor
    \GPcolortrue
  }{}%
  \@ifundefined{ifGPblacktext}{%
    \newif\ifGPblacktext
    \GPblacktexttrue
  }{}%
  \let\gplgaddtomacro\g@addto@macro
  \gdef\gplbacktext{}%
  \gdef\gplfronttext{}%
  \makeatother
  \ifGPblacktext
    \def\colorrgb#1{}%
    \def\colorgray#1{}%
  \else
    \ifGPcolor
      \def\colorrgb#1{\color[rgb]{#1}}%
      \def\colorgray#1{\color[gray]{#1}}%
      \expandafter\def\csname LTw\endcsname{\color{white}}%
      \expandafter\def\csname LTb\endcsname{\color{black}}%
      \expandafter\def\csname LTa\endcsname{\color{black}}%
      \expandafter\def\csname LT0\endcsname{\color[rgb]{1,0,0}}%
      \expandafter\def\csname LT1\endcsname{\color[rgb]{0,1,0}}%
      \expandafter\def\csname LT2\endcsname{\color[rgb]{0,0,1}}%
      \expandafter\def\csname LT3\endcsname{\color[rgb]{1,0,1}}%
      \expandafter\def\csname LT4\endcsname{\color[rgb]{0,1,1}}%
      \expandafter\def\csname LT5\endcsname{\color[rgb]{1,1,0}}%
      \expandafter\def\csname LT6\endcsname{\color[rgb]{0,0,0}}%
      \expandafter\def\csname LT7\endcsname{\color[rgb]{1,0.3,0}}%
      \expandafter\def\csname LT8\endcsname{\color[rgb]{0.5,0.5,0.5}}%
    \else
      \def\colorrgb#1{\color{black}}%
      \def\colorgray#1{\color[gray]{#1}}%
      \expandafter\def\csname LTw\endcsname{\color{white}}%
      \expandafter\def\csname LTb\endcsname{\color{black}}%
      \expandafter\def\csname LTa\endcsname{\color{black}}%
      \expandafter\def\csname LT0\endcsname{\color{black}}%
      \expandafter\def\csname LT1\endcsname{\color{black}}%
      \expandafter\def\csname LT2\endcsname{\color{black}}%
      \expandafter\def\csname LT3\endcsname{\color{black}}%
      \expandafter\def\csname LT4\endcsname{\color{black}}%
      \expandafter\def\csname LT5\endcsname{\color{black}}%
      \expandafter\def\csname LT6\endcsname{\color{black}}%
      \expandafter\def\csname LT7\endcsname{\color{black}}%
      \expandafter\def\csname LT8\endcsname{\color{black}}%
    \fi
  \fi
  \setlength{\unitlength}{0.0500bp}%
  \begin{picture}(4250.00,3400.00)%
    \gplgaddtomacro\gplbacktext{%
      \csname LTb\endcsname%
      \put(4048,525){\makebox(0,0)[l]{\strut{}\footnotesize $t$}}%
      \put(586,3136){\makebox(0,0)[l]{\strut{}\footnotesize $y$}}%
      \put(2587,421){\makebox(0,0)[l]{\strut{}\footnotesize $\tau_2$}}%
      \put(3176,421){\makebox(0,0)[l]{\strut{}\footnotesize $\tau_1$}}%
    }%
    \gplgaddtomacro\gplfronttext{%
    }%
    \gplbacktext
    \put(0,0){\includegraphics{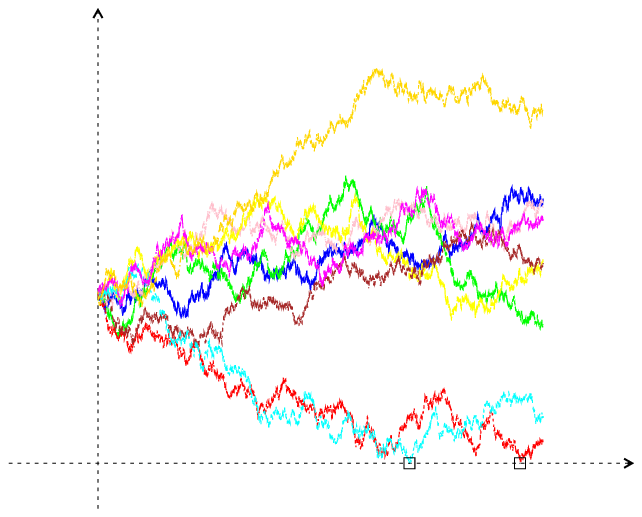}}%
    \gplfronttext
  \end{picture}%
\endgroup